\documentclass[prl,twocolumn,showpacs,preprintnumbers,amsmath,amssymb,superscriptaddress, bibnotes]{revtex4}

\usepackage{color}
\usepackage{graphicx}
\usepackage{dcolumn}
\usepackage{bm}

\begin{document}

\title{Superconductivity induced by longitudinal ferromagnetic fluctuations in UCoGe}

\author{T.~Hattori}
\email{t.hattori@scphys.kyoto-u.ac.jp}
\author{Y.~Ihara}
\altaffiliation{Present Address: Dept. of Phys., Grad. Sch of Sci, Hokkaido Univ., Sapporo 060-0810, Japan}
\affiliation{Department of Physics, Graduate School of Science, Kyoto University, Kyoto 606-8502, Japan}

\author{Y.~Nakai}
\altaffiliation{Present Address: Grad. Sch. of Sci, Tokyo Metropolitan Univ., Hachioji, Tokyo 192-0397,Japan}
\author{K.~Ishida}
\affiliation{Department of Physics, Graduate School of Science, Kyoto University, Kyoto 606-8502, Japan}
\affiliation{TRIP, JST, Sanban-cho bldg., 5, Sanban-cho, Chiyoda, Tokyo 102-0075, Japan}

\author{Y.~Tada}
\altaffiliation{Present Address: Inst. for Solid State Phys., Univ. of Tokyo, Kashiwa, Chiba 277-8581, Japan}
\author{S.~Fujimoto}
\author{N.~Kawakami}
\affiliation{Department of Physics, Graduate School of Science, Kyoto University, Kyoto 606-8502, Japan}

\author{E.~Osaki}
\author{K.~Deguchi}
\author{N.~K.~Sato}
\affiliation{Department of Physics, Graduate School of Science, Nagoya University, Nagoya 464-8602, Japan}

\author{I. Satoh}
\affiliation{Institute for Materials Research, Tohoku University, Sendai 980-8577 Japan}

\date{\today}

\begin{abstract}
From detailed angle-resolved NMR and Meissner measurements on a ferromagnetic (FM) superconductor UCoGe ($T_{\rm Curie} \sim 2.5 $ K and $T_{\rm SC} \sim 0.6$ K), we show that superconductivity in UCoGe is tightly coupled with longitudinal FM spin fluctuations along the $c$ axis. We found that magnetic fields along the $c$ axis ($H \parallel c$) strongly suppress the FM fluctuations and that the superconductivity is observed in the limited magnetic-field region where the longitudinal FM spin fluctuations are active. These results combined with model calculations strongly suggest that the longitudinal FM spin fluctuations tuned by $H \parallel c$ induce the unique spin-triplet superconductivity in UCoGe. This is the first clear example that FM fluctuations are intimately related with superconductivity.  
\end{abstract}


\abovecaptionskip=-8pt
\belowcaptionskip=-11pt

\maketitle


The discovery of superconductivity in ferromagnetic (FM) UGe$_2$ opened up a new paradigm of superconductivity\cite{SaxenaNature00, AokiNature01}, since most unconventional superconductivity has been discovered in the vicinity of an antiferromagnetic (AFM) phase\cite{MathurNature98}. 
From the theoretical point of view, in an itinerant FM superconductor with the presence of a large energy splitting between the majority and minority spin Fermi surfaces, exotic spin-triplet superconductivity is anticipated, in which pairing is between parallel spins within each spin Fermi surface. 
In addition, it has been argued that critical FM fluctuations near a quantum phase transition could mediate spin-triplet superconductivity\cite{FayPRB80}. However, there have been no experimental results indicating a relationship between FM fluctuations and superconductivity.

Among the FM superconductors discovered so far, UCoGe is one of the most readily explored experimentally, because of its high superconducting (SC) transition temperature ($T_{\rm SC}$) and low Curie temperature ($T_{\rm Curie}$) at ambient pressure\cite{HuyPRL07}. 
Microscopic measurements have shown that superconductivity occurs within the FM region, resulting in microscopic coexistence of ferromagnetism and superconductivity\cite{VisserPRL09, OhtaJPSJ10}. 
Studies of the SC upper critical field ($H_{\rm c2}$) and its angle dependence along each crystalline axis have reported remarkable enigmatic behavior\cite{HuyPRL08, AokiJPSJ09}: superconductivity survives far beyond the Pauli-limiting field along the $a$ and $b$ axes, whereas $H_{\rm c2}$ for fields along the $c$ direction ($H_{\rm c2}^c$) is as small as 0.5 T. Colossal $H_{\rm c2}$ for fields along the $a$ and $b$ axes seems to suggest spin triplet pairing. 
In addition, a steep angle dependence of $H_{\rm c2}$ was reported when the field was tilted slightly from the $a$ axis toward the $c$ axis\cite{AokiJPSJ09}. The observed characteristic $H_{\rm c2}$ behavior is one of mysterious features of SC UCoGe and its origin can be related to the mechanism of the superconductivity.

Unlike the three dimensional crystal structure, magnetic properties are strongly anisotropic\cite{HuyPRL08}. 
The magnetization has Ising-like anisotropy with the $c$ axis as a magnetic easy axis, and direction-dependent nuclear-spin lattice relaxation rate ($1/T_1$) measurements on a single crystalline sample have revealed the magnetic fluctuations in UCoGe to be Ising-type FM ones along the $c$ axis (longitudinal FM spin fluctuations)\cite{IharaPRL10}. 
Here, we report from precise angle-resolved $1/T_1$ and $H_{\rm c2}$ measurements how the longitudinal FM spin fluctuations are sensitively affected by the fields along the $c$ axis ($H^c$), and are linked with the superconductivity. In addition, with the aid of model calculations, we unveil the role of the FM fluctuations as pairing glue in this compound, concomitantly resolving the above-mentioned puzzle of $H_{\rm c2}$.

A single crystal of UCoGe was grown by the Czochralski pulling method in a tetra-arc furnace under high-purity argon. 
A $1.65 \times 1.65 \times 1.89$ mm$^3$ with a mass of 55.8 mg sample was cut by spark erosion from the single crystalline ingot. 
\begin{figure}[tb]
\begin{center}
\includegraphics[width=6cm,clip]{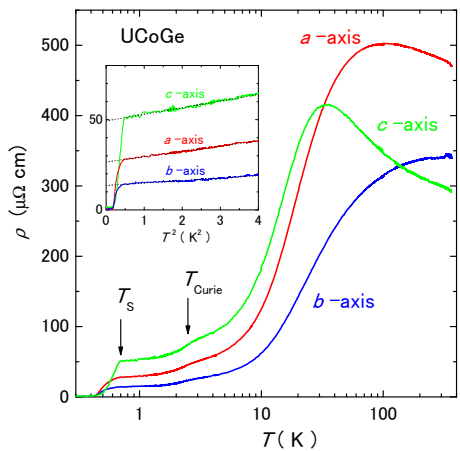}
\end{center}
\caption{(Color online) Temperature dependence of resistivity along each axis in the single-crystal UCoGe. The inset shows the plot of the resistivity against $T^2$ below 2 K. }
\label{Fig.1}
\end{figure}
The FM transition temperature $T_{\rm Curie}$ was evaluated to be $2.55 \pm 0.1$ K from the Arrot plots, and onset and midpoint SC transition temperatures were determined from ac susceptibility as 0.70 and 0.57 K, respectively. 
Clear anomalies in the specific heat were observed at $T_{\rm Curie}$ and $T_{\rm SC}$, confirming that two anomalies are the bulk transitions.  
Resistivity along each direction was measured and is shown in Fig.~1. 
The sample showed a large residual resistivity ratio ($RRR$) of approximately 30 along the $b$ axis. 
The temperature dependence below 2 K is approximately expressed as $\rho(T)=\rho_0+AT^2$  as shown in the inset. 
From the anisotropy of $A$ coefficient, the mass anisotropy is estimated as $m^*_c / m^*_b \sim 1.65$ using a relation of $m^* \propto \sqrt{A}$.
It is worth noting that a huge ratio of $H_{\rm c2}^a$ to $H_{\rm c2}^c$ ($H_{\rm c2}^a / H_{\rm c2}^c > 20$) cannot be explained by anisotropy of the conduction-electron mass. 

Low-energy magnetic fluctuations are sensitively probed by $1/T_1$ measurements.  
The single crystal was aligned such that external fields were applied within the $bc$ plane, and $\theta$ is defined as the angle between the applied field and the $b$ axis. 
We used a split-coil superconducting magnet with a single-axis rotator. 
NMR spectra for fields along three crystal axes and the locus of NMR peaks when field is rotated in the $ab$ and $bc$ plane were already shown in the literature\cite{THattori2011}.
When a nucleus with a spin larger than unity sits at a position where the electric field gradient (EFG) is finite, the nuclear quadrupole interaction splits the NMR spectrum.
Since the EFG parameters for Co site have determined from the previous $^{59}$Co NQR / NMR experiments\cite{OhtaJPSJ10, IharaPRL10} ( the quadrupole interaction of 2.85 MHz $\sim$ 0.28 T and the direction of maximum EFG principal axis of 10$^{\circ} $ from $a$ axis in $ac$ plane ), we can check $\theta$ from the quadrupole-split NMR spectrum.
The angle-dependent NMR spectra are well simulated by these EFG parameters\cite{THattori2011}.

When $1/T_1$ is measured in an external magnetic field far exceeding the electric quadrupole interaction ($\mu_0 H\gg 0.28$ T), the direction of the field is regarded as the quantization axis for nuclear spins. 
In these conditions, the nuclear spins are relaxed by transverse components of local hyperfine-field fluctuations at the nuclear site, which are produced by electron spins. 
Thus, $1/T_1$ measured in a field along the $\alpha$ direction is written in terms of fluctuating hyperfine fields along the $\beta$ and $\gamma$ directions as
\begin{equation}
\frac{1}{T_1^{\alpha}} \propto \left<\left(\delta H^{\beta}\right)^2 \right> + \left<\left(\delta H^{\gamma}\right)^2 \right>, 
\end{equation}
where the $\alpha$, $\beta$ and $\gamma$ directions are mutually orthogonal.  

The angle dependence of $1/T_1$ is measured at the central peak in each spectrum. 
The recovery curves $R$($t$) = 1- $m$($t$)/$m$($\infty$) of the nuclear magnetization $m$($t$), which is the nuclear magnetization at a time $t$ after a saturation pulse, can be fitted by the theoretical function for $I$ = 7/2 with a single component throughout the measured field and temperature range. 
Thus, the electronic state is considered to be homogeneous in the whole region, and reliable $1/T_1$ values were obtained. 

\begin{figure}[tb]
\begin{center}
\includegraphics[width=8.0cm,clip]{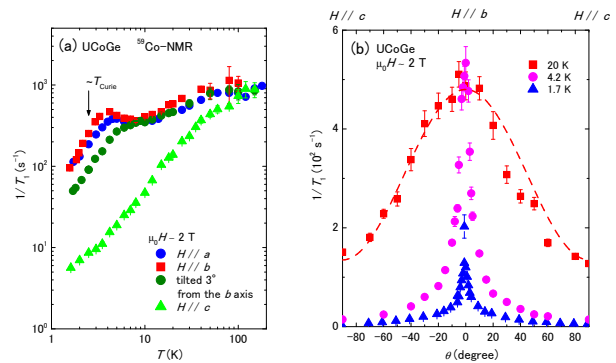}
\end{center}
\caption{(Color online) (a) Temperature dependence of $1/T_1$ measured in fields along the three crystal axes under $\sim$ 2 T, and tilted by 3 degrees from the $b$ axis. The broad peak in $1/T_1$, ascribed to the FM anomaly, is strongly suppressed when the magnetic field is tilted by 3 degrees. (b) Angle dependence of $1/T_1$ in the bc plane at 20, 4.2 and 1.7 K under $\sim 2$ T. The angle dependence at 20 K can be consistently explained by eqn.(2), while these at 4.2 and 1.7 K cannot, indicative of the strong suppression of $1/T_1$ by $H^{c}$.}
\label{Fig.1}
\end{figure}
As shown in Fig.~2 (a), $1/T_1$ along the $a$ and $b$ axes, which is much larger than $1/T_1$ along the $c$ axis due to the Ising-type anisotropy, shows a broad peak around 4 K due to the FM critical fluctuations. 
That the peak temperature is slightly higher than $T_{\rm Curie}$ is conceivably due to a slight misalignment, since the peak is suppressed and shifts to higher temperatures when the external field is inclined by 3$^{\circ}$ away from the $b$ axis. 
The anisotropy of $1/T_1$ is largest around the peak temperature and extremely sensitive to the field angle. 
Figure 2 (b) shows the angle dependence of $1/T_1$ ($1/T_1(\theta)$) measured at 20, 4.2, and 1.7 K. 
For magnetic fields in the $bc$ plane, $1/T_1(\theta)$ is expressed as,  
\begin{equation}
\frac{1}{T_1}(\theta)=\frac{1}{T_1^b}\cos^2{\theta}+\frac{1}{T_1^c}\sin^2{\theta}.
\end{equation}
This equation can fully explain the smooth variation at 20 K, but not the sharp angle dependence observed at 4.2 and 1.7 K, which shows a cusp centered at $\theta = 0^{\circ}$. 
The steep angle dependence of $1/T_1$ is characteristic feature of the FM fluctuations at low temperatures.

\begin{figure}[tb]
\begin{center}
\includegraphics[width=8.0cm,clip]{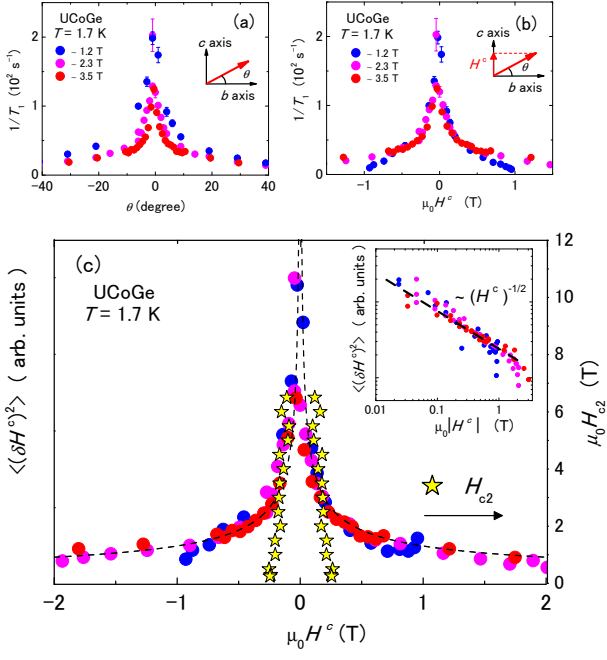}
\end{center}
\caption{(Color online) Angle dependence of $1/T_1$ in the $bc$ plane measured in three different magnetic fields at 1.7 K. (b) Plot of the $1/T_1$ against $H^c$. The $1/T_1$ data collapse onto a single curve when plotted against $H^c$. (c) $H^c$ dependence of magnetic fluctuations along the $c$ axis $\left<(\delta H^c)^2\right>$ at 1.7 K, extracted using eqn. (3). $H_{\rm c2}$ data shown in Fig. 3 is plotted against $H^c = H_{\rm c2}\sin{\theta}$. Inset: plot of $\left<(\delta H^c)^2\right>$ against $|H^c|$. The relation of $\left<(\delta Hc)^2\right> \propto 1/\sqrt{H^c}$ is shown by dashed lines in main panel (c). }
\label{Fig.3}
\end{figure}
To further examine this behavior, the angle dependence of $1/T_1$ was measured at 1.7 K under three different applied fields ($\mu_0 H \gg 0.28$ T). 
Although the cusp in $1/T_1$ versus angle becomes sharper with increasing $H$ as shown in Fig.~3 (a), the $1/T_1$ data collapse onto a single curve if $1/T_1$ is plotted against the $c$-axis component of $H$ ($H^c = H\sin{\theta}$) as seen in Fig.~3 (b). 
This unambiguously identifies $H^c$ as a relevant parameter for the longitudinal FM fluctuations. 
It is worth noting that the steep angle dependence of $1/T_1$ is observed even at 1.7 K $< T_{\rm Curie}$, indicating that the characteristic FM fluctuations survive in the FM ordered state. 
From the angle-resolved $1/T_1$ measurement, the $H^c$ dependence of the longitudinal FM fluctuations along the $c$ axis $\left<(\delta H^c)^2 \right>$ can be extracted from $1/T_1(\theta)$ and $1/T_1^c$ by combining equations (1) and (2) on the assumption that the magnetic fluctuations in the $ab$ plane are isotropic ($\left<(\delta H^a)^2\right> \sim \left<(\delta H^b)^2\right>$). 
Equations (1) and (2) then yield $\left<(\delta H^c)^2\right>$ as
\begin{equation}
\left<(\delta H^c)^2\right> \propto \frac{1}{\cos^2{\theta}}\left(\frac{1}{T_1}(\theta)-\frac{(1+\sin^2{\theta})}{2}\frac{1}{T_1^c}\right).
\end{equation}
Figure 3 (c) plots $\left<(\delta H^c)^2\right>$ at 1.7 K against $H^c$, exhibiting a strong suppression approximately as $H^{-1/2}$ as shown by the broken line in the inset. 
It should be emphasized that $\left<(\delta H^c)^2\right>$ at 20 K is independent of $H^c$ since the angle dependence of $1/T_1$ is well fitted by eqn.(2); the strong suppression of $\left<(\delta H^c)^2\right>$ by $H^c$ is observed only at low temperatures, where FM spin fluctuations are enhanced. 

\begin{figure}[tb]
\begin{center}
\includegraphics[width=8.0cm,clip]{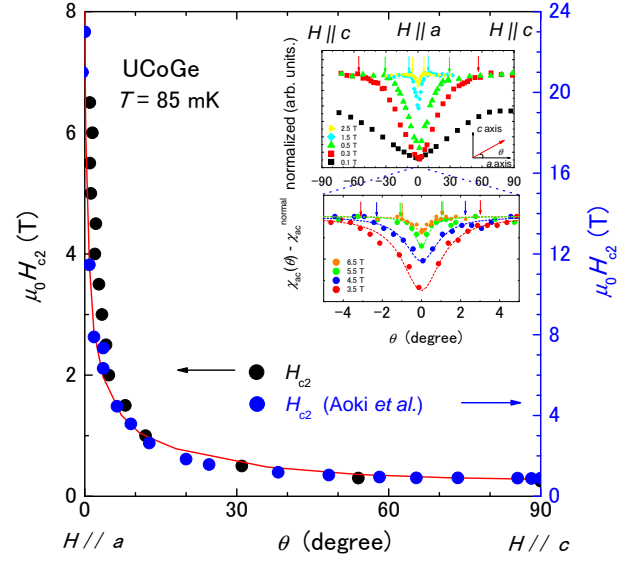}
\end{center}
\caption{(Color online) Angular dependence of $H_{\rm c2}$ in the $ac$ plane, which is determined by the onset of the Meissner signal. $H_{\rm c2}$ data reported by Aoki {\it et al} \cite{AokiJPSJ09}. are also plotted, which are determined by resistive measurements, and $H_{\rm c2}$ above 16 T was estimated from the linear extrapolation. The red curve is the calculation of the angle dependence of $H_{\rm c2}$ based on the spin-triplet A state by taking into account the field dependence of the Ising FM fluctuations shown in the inset of Fig. 3 (c). Inset: angular dependences of the Meissner signals measured in various magnetic fields. The arrows denote the onset angles, below which the Meissner signals come out. }
\label{Fig.4}
\end{figure}
A crucial question is whether such a characteristic feature of the FM fluctuations is related to superconductivity in UCoGe. 
To answer this question, we investigated the angle dependent SC Meissner signal at 85 mK with high frequency ac magnetic susceptibility measurements by observing the tuning frequency of the NMR circuit.
When the single-crystal UCoGe undergoes a SC phase transition, $\chi_{\rm bulk}$ becomes negative due to the Meissner effect and thus the frequency increases in the SC state.
Shift of the tuning frequency was monitored as a function of $\theta$. The angular dependence of $H_{\rm c2}$ in the $ac$ plane, which is determined from the onset of the Meissner signal is  shown in the inset of Fig.~4, since unusual enhancement of $H_{\rm c2}$ along the $b$ axis was reported in a field larger than 4 T\cite{AokiJPSJ09}. 
The $H_{\rm c2}$ enhancement is another interesting phenomenon in UCoGe but beyond the scope of this paper. 
The angular dependence of $H_{\rm c2}$ is qualitatively the same as previous report\cite{AokiJPSJ09}, but the values are different by a factor 3 from each other, as shown in the main panel of Fig.~4. 
As pointed out above, this steep angle dependence of $H_{\rm c2}$ cannot be explained at all with the anisotropic Ginzberg-Landau (GL) formula even if field dependence of $m^*$ is taken into account. 
In contrast, if $H_{\rm c2}$ is plotted against $H^c$ as shown in Fig.~3 (c), we notice that superconductivity is observed in this narrow field region where the longitudinal FM spin fluctuations are active. 
This coincidence strongly suggests that the longitudinal FM spin fluctuations play a crucial role in the superconductivity, and the strong suppression of $H_{\rm c2}$ along the $c$ axis is ascribed to the weakening of pairing interactions due to the suppression of the FM fluctuations by $H^c$.

This scenario is confirmed by theoretical calculations which successfully explain the above mentioned characteristic features of the SC properties. 
We consider an electron system with Ising FM fluctuations described by the susceptibility 
\begin{equation}
\chi_z(q, \Omega_n) \sim \left[\delta+q^2+|\Omega_n|/(\nu q)\right]^{-1}
\end{equation}
, where $\nu$ is approximately the Fermi velocity. 
Based on the results in Fig.~3 (c), we postulate that $\delta(H^c) \propto \left[1+c\sqrt{H^c/H^c_0} \right]$ with a magnetic field $H_0^c \sim 1$ T and a parameter $c \sim O(1)$. 
It is worth noting that this form for $\delta~(H^c)$ is quite different from the ordinary mean-field-like dependence, but that an anomalous $\sqrt{H^c}$ dependence is likely consequence of semi-metallic-like 5$f$-band structure in UCoGe. 
Two spin-triplet SC states coexisting with ferromagnetism have been proposed by Mineev, of the form $d \sim (a_1k_a+ia_2k_b, a_3k_b+ia_4k_a, 0)$ near the $\Gamma$ point corresponding to point node symmetry (A state) and $d \sim (b_1k_c+ib_2k_ak_bk_c, ib_3k_c+b_4k_ak_bk_c, 0)$ corresponding to horizontal line node symmetry (B state), where ${a_i}$ and ${b_i}$ are coefficients\cite{MineevPRB02}. 
We calculated angle dependence of $H_{\rm c2}$ by solving the Eliashberg equation for each of the SC states, and the results agree well with the experiments when we use $c = 5$ for the A state, as shown with the red curve in Fig.~4. 
 
\begin{figure}[tb]
\begin{center}
\includegraphics[width=7cm,clip]{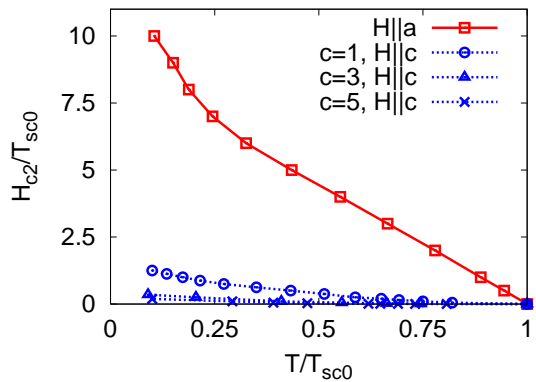}
\end{center}
\caption{(Color online) Temperature dependence of $H_{\rm c2}$ along the $a$ and $c$ axes calculated with the A state with the parameter of $c = 1, 3$ and 5.  }
\label{Fig.3}
\end{figure}
Temperature dependence of $H_{\rm c2}$ along the $a$ and $c$ axes was calculated based on the A state with $c = 1, 3$ and 5, and the results are shown in Fig.~5. 
It is found that the qualitative behaviors of $H_{\rm c2}$ are independent of the value of $c$. Interestingly, the similar anisotropic $H_{\rm c2}$ behavior with the enhancement of $H_{\rm c2}$ along the $a$ axis at low temperatures was actually observed\cite{AokiJPSJ09}.  
These agreements strongly support the scenario that the longitudinal FM spin fluctuations in UCoGe mediate the spin-triplet superconductivity with the A-state symmetry, since the A state can explain the experimental observation consistently, but the B state can not reproduce the experimental results. 
We stress that the unusually strong angular dependence of $H_{\rm c2}$ is extremely difficult to be explained with a mean-field-like $\delta(H^c)$-dependence, but may be readily modelled with that deduced from the present experiment.

In conclusion, we demonstrate that a huge anistropic ratio of $H_{\rm c2}$ cannot be explained by the anisotropic GL model, since conductivity of UCoGe is rather 3D. 
We found from the angle-resolved NMR measurements that the magnetic fields along the $c$ axis $H^c$ strongly suppress the critical FM fluctuations with Ising anisotropy, and that their suppression is intimately related to the unusual anisotropic properties of the superconductivity. 
The theoretical calculations based on the experimentally obtained $\delta(H^c)$ dependence explain the strong suppression of $H_{\rm c2}$ along the $c$ axis. 
Our experimental results and theoretical analysis strongly suggest that the critical FM fluctuations with Ising anisotropy mediate the spin-triplet superconductivity in UCoGe. 

The authors thank D. C. Peets, S. Yonezawa, and Y. Maeno for experimental support and valuable discussions, and D. Aoki, J. Flouquet, A. de Visser, A. Huxley, H. Harima, and H. Ikeda for valuable discussions. 
This work was partially supported by Kyoto Univ. LTM centre, Yukawa Institute, the "Heavy Electrons" Grant-in-Aid for Scientific Research on Innovative Areas  (No. 20102006, No. 21102510,  No. 20102008, and No. 23102714) from The Ministry of Education, Culture, Sports, Science, and Technology (MEXT) of Japan, a Grant-in-Aid for the Global COE Program ``The Next Generation of Physics, Spun from Universality and Emergence'' from MEXT of Japan, a grant-in-aid for Scientific Research from Japan Society for Promotion of Science (JSPS), KAKENHI (S) (No. 20224015) and FIRST program from JSPS .


\begin{thebibliography}{99}

\bibitem{SaxenaNature00}
S. S. Saxena, P. Agrwal, A. Ahilan, F. M. Grosche, R. K. W Haselwimmer, M. J. Steiner, E. Pugh, I. R. Walker, S. R. Julian, P. Monthoux, G. G. Lonzarich, A. Huxley, I. Sheiken, D. Braithwaite, and J. Flouquet,  Nature {\bf 406}, 587 (2000).

\bibitem{AokiNature01}
D. Aoki, A. Huxley, E. Ressouche, D. Braithwaite, J. Flouquet, J-P. Brison, E. Lhotel and C. Paulsen, Nature {\bf 413}, 613 (2001).

\bibitem{MathurNature98}
N.~D.~Mathur, F.~M.~Grosche, S.~R.~Julian, I.~R.~Walker, D.~M.~Freye, R.~K.~W.~Haselwimmer and G.~G.~Lonzarich, Nature {\bf 394}, 39 (1998).

\bibitem{FayPRB80}
D. Fay and J. Appel, Phys. Rev. B {\bf 22}, 3173 (1980).

\bibitem{HuyPRL07}
N. T. Huy, A. Gasparini, D. E. de Nijs, Y. Huang, J. C. P. Klaasse, T. Gortenmulder, A. de Visser, A. Hamann, T. G\"{o}rlach and H. v. L\"{o}hneysen, Phys. Rev. Lett. {\bf 99}, 067006 (2007).

\bibitem{VisserPRL09}
A. de Visser, N. T. Huy, A. Gasparini, D. E. de Nijs, D. Andreica, C. Baines and A. Amato: Phys. Rev. Lett. {\bf 102}, 167003 (2009).

\bibitem{OhtaJPSJ10}
T. Ohta, T. Hattori, K. Ishida, Y. Nakai, E. Osaki, K. Deguchi, N. K. Sato, and I. Satoh, J. Phys. Soc. Jpn. {\bf 79}, 023707 (2010).

\bibitem{HuyPRL08}
N. T. Huy, D. E. de Nijs, Y. K. Huang, and A. de Visser, Phys. Rev. Lett. {\bf 100}, 077002 (2008).

\bibitem{AokiJPSJ09}
D. Aoki, T. D. Matsuda, V. Taufour, E. Hassinger, G. Knebel, and J. Flouquet, J. Phys. Soc. Jpn. {\bf 78}, 113709 (2009).

\bibitem{IharaPRL10}
Y. Ihara, T. Hattori, K. Ishida, Y. Nakai, E. Osaki, K. Deguchi, N. K. Sato, and I. Satoh, Phys. Rev. Lett. {\bf 105}, 206403 (2010).

\bibitem{THattori2011}
T. Hattori, Y. Ihara, K. Ishida, Y. Nakai, E. Osaki, K. Deguchi, N. K. Sato, and I. Satoh, J. Phys. Soc. Jpn. {\bf 80}, SA007 (2011).

\bibitem{MineevPRB02}
V. P. Mineev, Phys. Rev. B {\bf 66}, 134504 (2002).

\end{thebibliography}
\end{document}